# On Demand Data Analysis and Filtering for Inaccurate Flight Trajectories

Massimiliano Zanin, David Perez
The Innaxis Research Institute, Spain
{mzanin, dp}@innaxis.org

Kumardev Chatterjee
Thales Group, Belgium
kumardev.chatterjee@thalesgroup.com

Dimitrios S. Kolovos, Richard F. Paige
Department of Computer Science
University of York, UK
{dkolovos, paige}@cs.york.ac.uk

Andreas Horst, Bernhard Rumpe
Department of Computer Science
RWTH Aachen University, Germany
{horst, rumpe}@se-rwth.de

*Abstract*—This paper reports on work performed in the context of the COMPASS SESAR-JU WP-E project, on developing an approach for identifying and filtering inaccurate trajectories (*ghost flights*) in historical data originating from the EUROCONTROL-operated Demand Data Repository (DDR).

*Foreword*—This paper describes a project that is part of SESAR Workpackage E, which is addressing long-term and innovative research. The project was started early 2011 so this description is limited to an outline of the project objectives augmented by some early findings.

## I. Introduction

COMPASS is a collaborative project funded in the context of WP-E of the SESAR-JU initiative. The aim of the project is to investigate the application of techniques such as data mining, complex event processing, domain-specific modelling and root-cause analysis for the identification of safety-related event patterns in the operation of Air Traffic Management (ATM) systems. The identified patterns will be used for the development of a new system that will detect their occurrences in live data and produce early warnings about situations that – if left to develop – could compromise the safe operation of an Air Traffic Management (ATM) system.

To identify safety-related patterns which can later on be detected in the early safety-warning system (ESWS), past data can be used for mining and analysis. In this paper we present results of our work on analysing and filtering data from the Demand Data Repository (DDR), a EUROCONTROL repository containing historical information for 4D trajectories of flights crossing the European air space. More specifically, we report on limitations of the DDR and demonstrate an approach for filtering out inaccurate flight trajectories (*ghost flights*).

The remainder of the paper is organised as follows. Section II provides background and motivation for the problem and its context. Section III provides an overview of DDR, discusses the DDR data considered in this work and the identified limitations – most notably their low temporal and spatial resolution and the existence of *ghost* flights. Section IV discusses the concept of *ghost* flights in more detail and presents an approach for identifying and eliminating them from the DDR data. Section V summarises the findings and concludes the paper.

## II. Background

Air Traffic Management systems are predominantly characterised by complexity and heterogeneity; they encompass a sophisticated ecosystem of heterogeneous systems and procedures that ensure the excellent safety records of the aviation industry both on the ground and in the air. ATMs are predominantly safety-related or safety-critical systems, since failure in functionality can result in accidents, loss of property, or loss of life. During the operation of an ATM system, its components and systems produce a high volume of system events (e.g. health status of individual devices, temperature/proximity readings), and perform measurements (e.g. volume of traffic, structural complexity of the airspace). Of particular interest to safety management are combinations of events and measurements that can lead to scenarios where the safe operation of the system is compromised. These scenarios include situations where uncertainty is high, e.g., weather conditions that increase the uncertainty of the forecasted flight path, or where there are a number of complex events processed. In this context, a major challenge is to identify such hazardous scenarios as they occur and issue warnings to human operators sufficiently early, so that they can take preventative or mitigating actions.

The high volume of potentially interrelated events produced across the system, and the large number of possible combinations of events that can compromise safety render manual monitoring and management extremely challenging. The need for automated mechanisms that can filter and organise high volumes of heterogeneous, incomplete or unreliable information in an intelligent manner is imperative.

The aim of COMPASS is to combine state-of-the-art technology from the fields of complexity science, data-mining, intelligent modelling and complex event processing to enable engineers to mine previously unknown patterns from past data. It will allow engineers to filter and enrich these patterns through their expertise using intelligent domain-specific modelling tools, and then use these patterns to automatically monitor the running system in order to identify and report on situations which can compromise the safe operation of the ATM system.

To achieve this goal, COMPASS concentrates on three interrelated objectives, which form the basis of its three main technical work packages. The first objective is to identify crude patterns of events that have historically led to loss of safety in ATM systems. This shall be achieved by performing data mining on large-scale databases containing events related to incidents where safety was compromised in ATMs in the past. These crude patterns will then be refined by domain experts using a tailored domain-specific modelling language and supporting intelligent modelling tools. Finally, the patterns defined by domain experts will be used at runtime to monitor the events produced by the components and sub-systems of an operating ATM system in order to identify developing pattern matches and issue early warnings to human operators.

### A. Expected Results and Benefits to the ATM community

The project will deliver new safety-related scientific and technical ATM applications and techniques that go beyond the nominal SESAR timescales. It will deliver novel automated safety warning technology that offers promise in reducing the amount of human intervention in identifying hazardous situations and generating warnings to ATM experts. In particular, it will exploit and bring within the scope of SESAR research and development several novel technologies and theories from the ICT community tailored for the ATM domain so that they are easily exploited by ATM engineers, particularly model-driven engineering, complex event processing and automated safety warning analysis. Proof of concepts will be delivered that offer exploitation in the interim period, through use of standardised ICT technology such as approaches to intelligent systems modelling, and complex event processing. The CEP and RCA techniques developed will contribute to the ATM community by furthering its understanding of emergent behaviour demonstrated by complex ATM systems and allow ATM engineers to reason about it at an appropriate level of abstraction. The automated early safety warning mechanisms proposed will provide warnings early enough that preventive/corrective measures can be taken to ensure the safe operation of the ATM system of systems by degrading its provided services in a graceful manner. Finally, all tasks performed within the project shall inherently assume non-determinism (incompleteness, incorrectness, unreliability) in the involved data addressing a crucial area of concern within the ATM community. All combined, these techniques offer the promise of allowing us to increase the amount of automation in the ATM domain. We see substantial benefits for SESAR and the ATM domain in adopting and attempting to exploit concepts, technologies and theories from these specific ICT contributions. In the medium to long term we argue that these novel concepts should be exploited within SESAR and that they hold substantial promise in making a step-change in ATM.

### III. Historical Data Analysis

The idea of post-analysing historical ATM data to identify patterns related to safety is not new. Nazeri et al. analysed global metrics to forecast which air sectors were most incident-prone [1], and to understand the relation between these global metrics and the occurrence of incidents [2]. In [3], the spatiotemporal distribution of safety related events provided by the EUROCONTROL Automatic Safety Monitoring Tool was analysed. In [4] safety events were studied from a Complexity Science point of view, by creating a Complex Network and studying its topological characteristics. In this section, we present the results of our work on analysing data from the Demand Data Repository (DDR).

DDR is a repository developed by EUROCONTROL inside the DMEAN programme, which allows users to access historical and forecasted air traffic demand, with the main aim of supporting more efficient operation planning. The repository contains complete historical information for 4D trajectories of all flights crossing the European air space, built on Central Flow Management Unit (CFMU) data, and information related to the configuration of airspaces and airports (capacities, runway configurations etc.).

Flight trajectories are codified and stored in *segments*. For each one of these segments, several pieces of information are available: general information about the flight such as departure and destination airports, aircraft type, and flight number; coordinates of the initial and final points of the segment codified by unique identifiers, and by standard 4D coordinates; phase of the flight, and distance covered. Furthermore, information is organised into two different datasets [5]; for each day of operation, a *m1* file contains the trajectories as reported in the last filed flight plan, while a *m3* file includes corrected information for flights where excessive vertical, horizontal or time variation compared to flight plan was detected by radar.

The DDR dataset analysed in the first part of the COMPASS project extends from the 1st to the 28th of June 2011. The evolution of the number of segments according to the day is reported in Fig. 1 (Top), both for the *m1* and *m3* datasets. We observe that there is an important difference – of the order of 5% – in the number of segments corresponding to the planned and executed flights: this is due to flights that have been cancelled, but especially to aircraft that have received direct vectors towards their destination, and therefore have not crossed several intermediate points. This is further confirmed by Fig. 1 (Bottom), which illustrates the percentage of segments, in the *m3* dataset, that do not correspond with a segment in the *m1* file.

One of the most important requirements that have been identified within COMPASS for the historical radar tracks is high spatial and temporal resolution of the trajectories. The reason for this is that with a low temporal resolution, it would not be feasible to understand if the behaviour of an aircraft (for instance, a level change) has been performed with enough time to ensure a good separation with other traffic in the area, or has been the result of a last-second decision. Thus, it is important to assess the length and duration of these segments, as they are directly related with the quantity of information that has been lost.

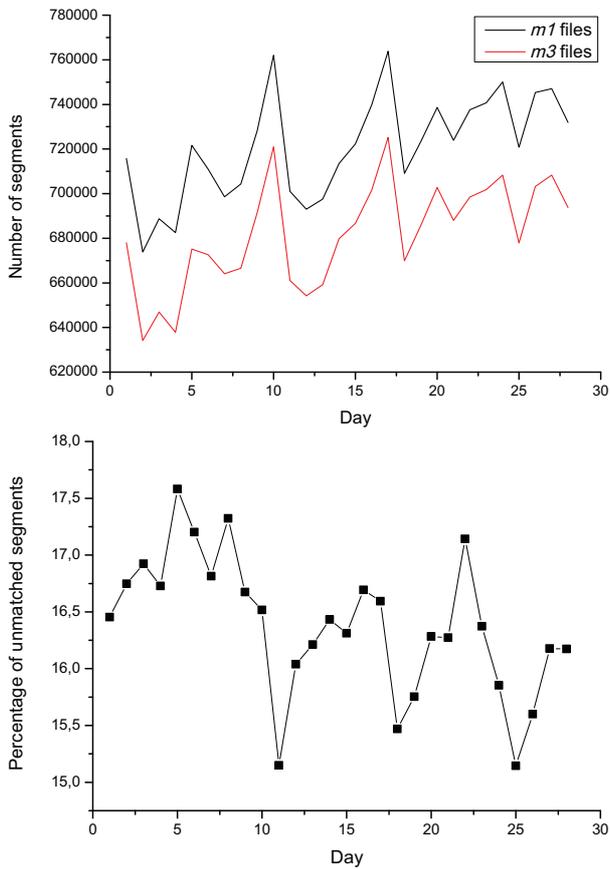

Fig. 1.  (Top) Representation of the number of segments, for the month of June 2011, recorded in the *m1* and *m3* files of the DDR dataset. (Bottom) Percentage of segments of the *m3* files that were not present in the *m1* dataset; this mismatch represents situations where there has been a change with respect to the planned route.

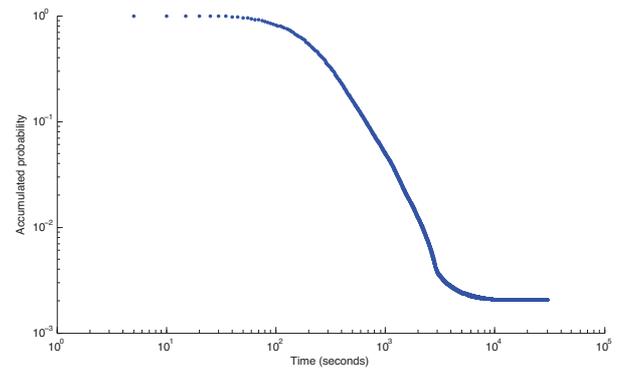

Fig. 2.  Histogram of the duration of segments (in seconds), in a *log-log* scale.

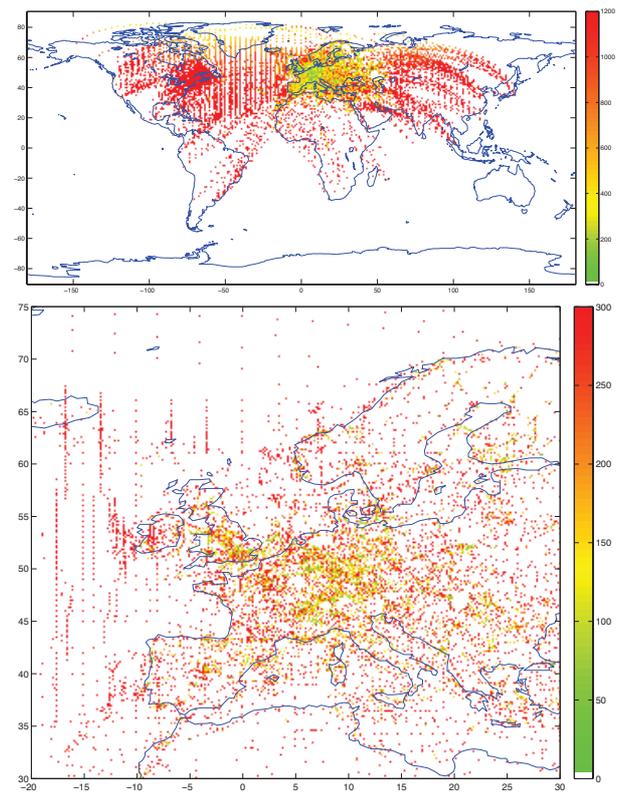

Fig. 3.  Representation of the number of segments, for the month of June 2011, recorded in the *m1* and *m3* files of the DDR dataset.

Fig. 2 represents the distribution of the temporal duration (in seconds) of segments. While 80% of segments have a duration of less than 120 seconds, it is important to notice the long tail of the distribution, that is, the significant number of segments covering more than 10 minutes.

This distribution is global, in the sense that it does not take into account the geographical position of segments. Clearly, this aspect can be extremely relevant; if, for instance, all long segments are located in specific areas that are not of interest for the study (as, for instance, Atlantic routes), the effective duration of relevant segments will be lower than expected. In Fig. 3, the spatial position of segments has been included in the analysis; the colour of each point represents the mean duration of segments beginning within a square centred at that point, and of size $1° \times 1°$ (for Fig. 3 (Bottom), each cell in the grid has a dimension of $0.1° \times 0.1°$).

Although trajectories inside Europe are sampled at a higher resolution, it is important to notice that only a few points are green, that is, few segments have a duration of less than a minute. This is the first problem that has been encountered within this project: the use of the DDR dataset of aircraft trajectories does not provide the best possible source of information for a reliable safety pattern mining process.

## IV. IDENTIFICATION OF *ghost* FLIGHTS

As a first step towards the identification of safety-related patterns in the historical ATM data available, some simple algorithms for detecting conflicts, i.e., situations involving a loss of separation minima, have been developed and ran against one day of operations. The simplest one was aimed at detecting all pairs of aircraft crossing the same significant point, at the same flight level, with a separation of less than 2 minutes; furthermore, the altitude at which both aircraft have crossed the same point should have been of at least 20.000 fts, with both aircraft in the en-route phase: these conditions were included in order to eliminate events in approach or departure phases.

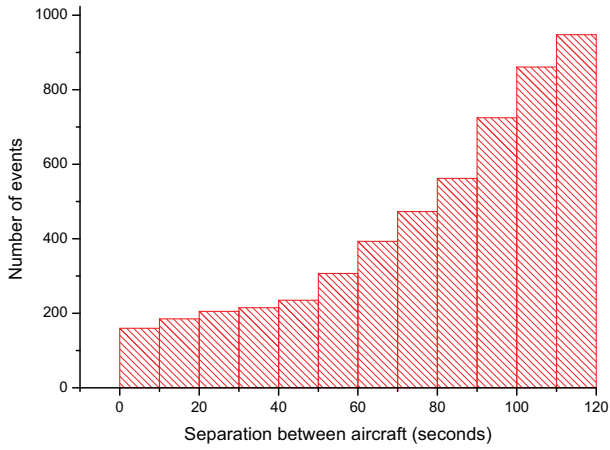

Fig. 4. Cumulative number of conflicts detected in one day of operation, as a function of the time separation between the two aircraft.

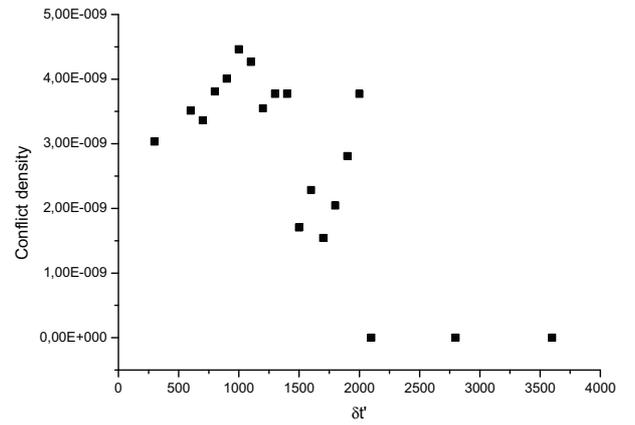

Fig. 5. Representation of the density of conflicts (defined as the number of conflicts detected in the dataset, scaled by the square of the number of traffics) as a function of $\delta t'$. Note that a clear phase transition appears for $\delta t' \approx 2000s$.

5274 pairs of segments fulfilled these conditions; the distribution of the separation between aircrafts, in seconds, is represented in Fig. 4. This value is, clearly, abnormally high, especially if compared to the 1356 events reported to EU-ROCONTROL in 2009 [6]; but, more important, it reveals that the information included in the dataset is not completely trustworthy.

The reason for this mismatch can be found in the definition of the content of both datasets. As already explained, the *m3* file combines out-of-date information of flight plans, with updated information originating from radar measurements when the difference between the real and planned situation of an aircraft exceeds some given thresholds. Therefore, the trajectories corresponding to many flights (i.e., of those flights which have followed the original flight plan) are not correctly updated, and many unresolved conflicts appear.

Once this problem was identified, a strategy was elaborated aiming at identifying (and, eventually, delete from the dataset) those flights that were not correctly updated (*ghost* flights).

If segments of flights are updated when the planned and executed trajectories differ more than a given threshold, it should be possible to detect that threshold; after that, all segments that correspond to a difference (between the information stored in the *m1* and *m3* files) lower than such thresholds are marked as *ghost* segments, and eventually excluded from the analysis. In order to simplify the analysis, only a temporal threshold $\delta t$ is considered, thus disregarding a potential spatial threshold.

It should be noticed that the actual value of the threshold $\delta t$ is not known; nevertheless, in principle, it is possible to detect such a value, as it must be associated to a phase transition in the dynamics of the system. Let us suppose that the value of such threshold is fixed to an arbitrary number $\delta t'$, and that flights that have suffered from delays smaller than $\delta t'$ are deleted from the dataset; in other words, we just consider the traffic composed by segments fulfilling the following condition:

$$abs(T_{m_1} - T_{m_3}) > \delta t' \qquad (1)$$

where $T_{m_1}$ and $T_{m_3}$ are the time at which one aircraft has crossed a given significant point, according to the *m1* and the *m3* files respectively. Once the dataset has been filtered, it is possible to detect the number of losses of separations, or, more specifically, the density of such events:

$$D(\delta t') = \frac{n_{LoS}(\delta t')}{s(\delta t')^2} \qquad (2)$$

where $n_{LoS}$ is the number of events identified, and $s$ is the number of segments in the filtered dataset; notice how both values are defined as a function of $\delta t'$, i.e., of the threshold used in the filtering process.

The evolution of such conflict density as a function of the threshold $\delta t'$ is represented in Fig. 5. Two different regimes can be found. First, on the right part, there is a region where no conflicts are detected; on the contrary, on the left side of the graph, there is a region with a high number of loss of separation events. From this, we can make the hypothesis that the transition from one regime to the other corresponds to the value of $\delta t'$ actually used in the data updating process, i.e., $\delta t = 2000s$.

In Fig. 6 is plotted the number of conflicts detected, when a threshold of $\delta t = 2000s$ is used to filter unreliable segments. Clearly, the result is more in line with the real situation, and the number of high-risk conflict has been reduced in two orders of magnitude.

## V. CONCLUSIONS

In this paper we have reported on work-in-progress on analysing and filtering DDR data in the context of COMPASS.

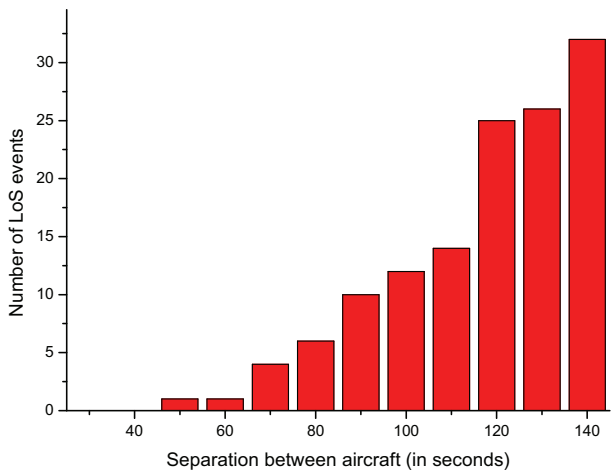

Fig. 6. Cumulative number of conflicts detected in one day of operation, as a function of the time separation between the two aircraft, after the deletion of *ghost* flights detected by the algorithm described in this contribution.

Our results have demonstrated that before they can be used for any meaningful analysis, DDR data need to be filtered in order to remove *ghost* flights and also need to be augmented with additional data-sources in order to improve their spatial and temporal resolution.